\documentclass[shortnote]{jpsj2} 
\title{Enhancement of Electronic Anomalies in Iron-Substituted La$_{2-x}$Sr$_x$Cu$_{1-y}$Fe$_y$O$_4$ around $x=0.22$}

\author{Tadashi \textsc{Adachi}\thanks{E-mail address: adachi@teion.apph.tohoku.ac.jp}, Hidetaka \textsc{Sato}, Masaki \textsc{Fujita}$^{1}$, Kazuyoshi \textsc{Yamada}$^{1}$, and Yoji \textsc{Koike}}

\inst{Department of Applied Physics, Graduate School of Engineering, Tohoku University, 6-6-05 Aoba, Aramaki, Aoba-ku, Sendai 980-8579 \\
$^{1}$Institute for Materials Research, Tohoku University, 2-1-1 Katahira, Aoba-ku, Sendai 980-8577} 

\kword{iron-substitution effect, La$_{2-x}$Sr$_x$CuO$_4$, electrical resistivity, magnetic susceptibility, overdoped regime}

\begin{document}
\maketitle

Since the early stage of the research on high-$T_{\rm c}$ superconductivity, much attention has been paid to the physics in the overdoped regime, e.g., the disappearance of the so-called pseudogap at a hole concentration per Cu, $p$, of $\sim 0.19$,~\cite{tallon} a topological change of the Fermi surface,~\cite{ino} a possible phase separation into superconducting (SC) and normal-state regions.~\cite{uemura,tanabe} 
In particular, around $x=0.22$ in the overdoped regime of La$_{2-x}$Sr$_x$CuO$_4$ (LSCO), significant anomalies have been observed thus far: a slight depression of the SC transition temperature $T_{\rm c}$,~\cite{kaki,tarou} the development of the Cu-spin correlation at low temperatures,~\cite{watanabe} the enhancement of lattice instability by the application of a magnetic field,~\cite{suzuki} and an anomalous insensitivity of the thermal conductivity to a magnetic field.~\cite{haidar} 
These are suggestive of a possible development of the stripe correlations of holes and spins~\cite{tranquada} around $x=0.22$, because these are analogous to those observed around $x=1/8$ where the stripe correlations are much developed. 

Recently, Fujita {\it et al}. have revealed from their elastic neutron-scattering experiment that incommensurate magnetic peaks around (1/2, 1/2) in the reciprocal lattice unit are enhanced through the partial substitution of Fe for Cu in LSCO around $p=1/8$.~\cite{fujita} 
Moreover, incommensurate elastic nuclear peaks around (0, 2) in the reciprocal lattice unit are observed in Fe-substituted samples around $p=1/8$, while they are unobservable in an Fe-free sample of $x=0.12$.~\cite{kimura} 
It is well known that superconductivity is strongly suppressed through Fe substitution in the whole SC regime.~\cite{xiao,arai}
Therefore, Fujita {\it et al}.~\cite{fujita} have suggested that Fe substitution is effective for the stabilization of the charge-spin stripe order around $p=1/8$. 

In this study, we have investigated Fe-substitution effects on electronic properties around $x=0.22$ in the overdoped regime of LSCO by measuring the electrical resistivity $\rho$ and the magnetic susceptibility $\chi$.

Polycrystalline samples of 1 \% Fe-substituted La$_{2-x}$Sr$_x$Cu$_{1-y}$Fe$_y$O$_4$ with $y=0.01$ were prepared by the ordinary solid-state reaction method. 
The details have been reported in our previous paper.~\cite{kaki}
All the samples were checked by powder X-ray diffraction analysis to be of the single phase. 
$\rho$ was measured by the standard dc four-probe method. 
$\chi$ measurement was carried out at a magnetic field of 10 Oe on both zero-field cooling and field cooling at low temperatures down to 2 K, using a superconducting quantum interference device magnetometer (Quantum Design, MPMS).

Figure 1 shows the temperature dependences of $\rho$ for typical values of $x$ in La$_{2-x}$Sr$_x$Cu$_{1-y}$Fe$_y$O$_4$ with $y=0$~\cite{kaki} and $y=0.01$. 
For $y=0$, the data except for $x=0.22$ exhibit a metallic behavior in the normal state, which is typical of overdoped high-$T_{\rm c}$ cuprates. 
However, a less metallic behavior is observed for $x=0.22$. 
For $y=0.01$, on the other hand, a weakly localized behavior is observed just above the onset $T_{\rm c}$ for $x=0.20$, 0.215, 0.225 and 0.235. 
This may be due to the scattering of holes by the large magnetic moment of Fe$^{3+}$ (spin quantum number $S=5/2$), because the behavior is unobservable for nonmagnetic Zn-substituted samples in the overdoped regime.~\cite{kaki,fukuzumi} 
For $x=0.2175$ and 0.22, unexpectedly surprisingly, it is found that the values of $\rho$ are larger than those of the other samples at room temperature and that a pronounced upturn is observed below $\sim 100$ K. 
Furthermore, the $T_{\rm c}$'s of $x=0.2175$ and 0.22 are lower than those of the other samples, which is clearly seen in Fig. 3, which will be mentioned later. 
These anomalous behaviors of $\rho$ are reminiscent of those observed around $x=0.22$ in Zn- and Ga-substituted LSCO.~\cite{kaki,koike}

Figure 2 shows the temperature dependence of $\chi$ on both zero-field cooling and field cooling. 
It is found that the onset $T_{\rm c}$ where the diamagnetism starts to appear is anomalously depressed for $x=0.220$ and 0.2225, around which anomalous behaviors of $\rho$ are observed, as shown in Fig. 1. 
These indicate that the anomalous decrease in $T_{\rm c}$ around $x=0.22$ is a bulk property of the samples. 

Figure 3 shows the Sr concentration $x$ dependence of $T_{\rm c}$ estimated from the $\rho$ and $\chi$ measurements, together with the data of Fe-free samples.~\cite{kaki} 
The definition of $T_{\rm c}$ has been mentioned in the caption of Fig. 3.
The values of $T_{\rm c}$ for $y=0.01$ are reduced by Fe substitution, which is consistent with the results of previous studies.~\cite{xiao,arai}
It is found that the slight depression of $T_{\rm c}$ observed for the Fe-free samples around $x=0.22$ becomes marked by Fe substitution. 
This resulting marked depression of $T_{\rm c}$ by Fe substitution reminds us of the Zn-substitution effect on the depression of $T_{\rm c}$ around $x=1/8$ in LSCO.~\cite{koike-zn}

The present results of the anomalous behaviors of $\rho$ and the marked depression of $T_{\rm c}$ around $x=0.22$ in Fe-substituted LSCO is qualitatively similar to those observed around $x=0.22$ in Zn-substituted LSCO,~\cite{kaki} where it has been suggested that dynamical stripe correlations are pinned and stabilized by Zn. 
Therefore, taking into consideration the Fe-substitution effects around $p=1/8$ in LSCO,~\cite{fujita} the present results strongly suggest that dynamical stripe correlations are stabilized by Fe around $x=0.22$, leading to a less metallic behavior of $\rho$ and to the suppression of superconductivity. 
Moreover, the depression of $T_{\rm c}$ induced by Fe substitution is more marked than that induced by Zn substitution~\cite{kaki} around $x=0.22$, suggesting that the large magnetic moment of Fe$^{3+}$ ($S=5/2$) is more effective for the stabilization of the stripe order than nonmagnetic Zn$^{2+}$ ($S=0$). 
Here, note that $x$ where $T_{\rm c}$ is depressed is not affected by Fe substitution, while it shifts to the low-$x$ side with Zn substitution.~\cite{kaki}
It is believed that Fe is substituted as Fe$^{3+}$ for Cu$^{2+}$, indicating a decrease in $p$ induced by Fe substitution. 
Therefore, it appears that the dip of $T_{\rm c}$ tends to shift to the low-doping side of $p\sim0.21$, reminiscent of the shift of the dip induced by Zn substitution around $x\sim0.22$.~\cite{kaki} 
A forthcoming study will determine whether or not stripe correlations develop around $x=0.22$ for Fe-substituted samples, which will soon be under way using neutron-scattering and muon-spin-relaxation experiments.

In summary, we have measured the temperature dependences of $\rho$ and $\chi$ for Fe-substituted LSCO in the overdoped regime, in order to investigate Fe-substitution effects on electronic properties around $x=0.22$. 
From the $\rho$ measurements, it has been found around $x=0.22$ that the values of $\rho$ are large at room temperature and that $\rho$ exhibits a pronounced upturn at low temperatures. 
Moreover, from the $\rho$ and $\chi$ measurements, it has been found that $T_{\rm c}$ is anomalously depressed around $x=0.22$. 
These results indicate that the electronic anomalies around $x=0.22$ are enhanced by Fe substitution, which might be related to the development of stripe correlations by Fe substitution.

\begin{figure}[tbp]
\begin{center}
\includegraphics[width=1.0\linewidth]{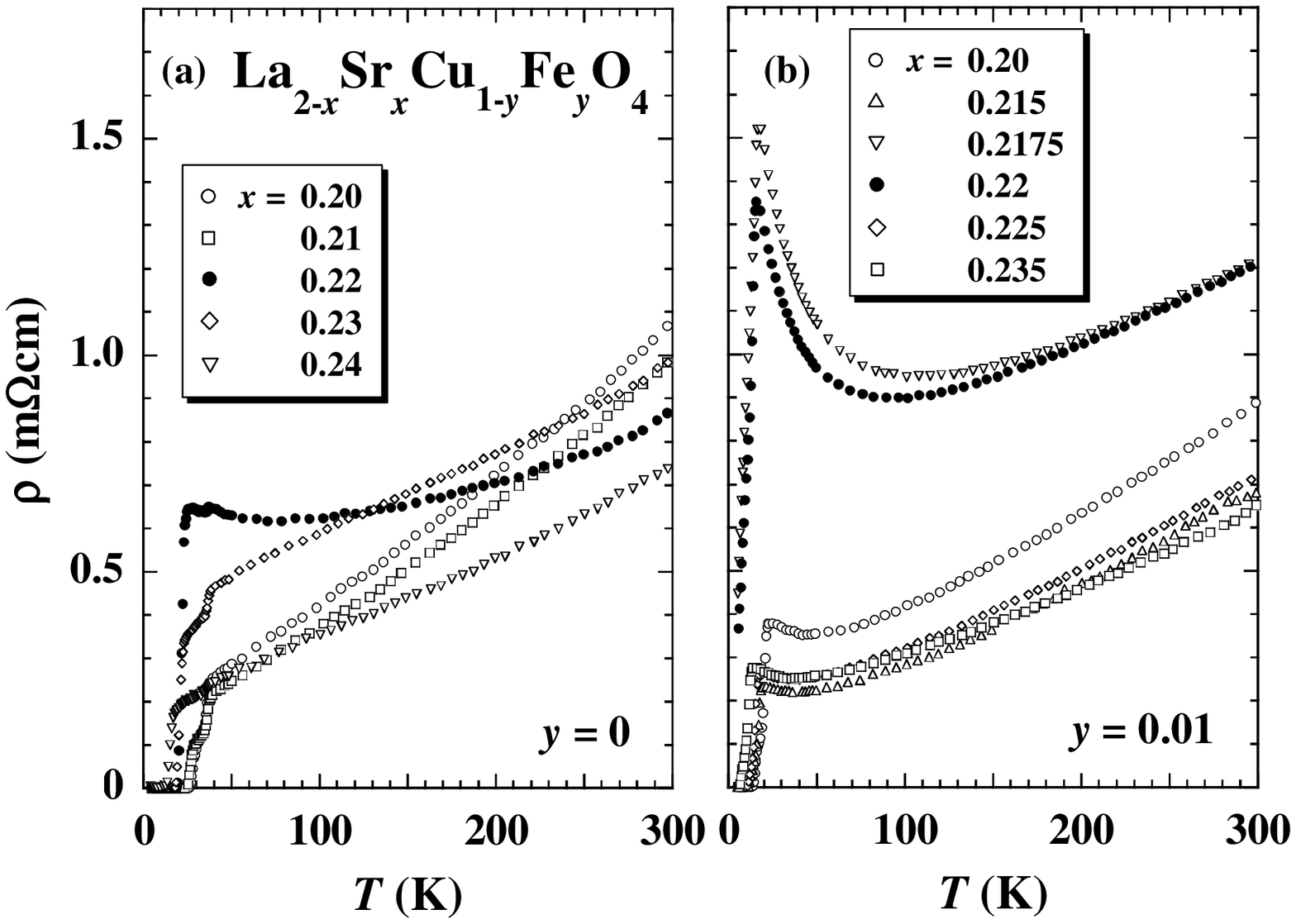}
\end{center}
\caption{Temperature dependences of the electrical resistivity $\rho$ for typical values of $x$ in La$_{2-x}$Sr$_x$Cu$_{1-y}$Fe$_y$O$_4$ with (a) $y=0$~\cite{kaki} and (b) $y=0.01$.}
\label{fig1} 
\end{figure}

\begin{figure}[tbp]
\begin{center}
\includegraphics[width=0.8\linewidth]{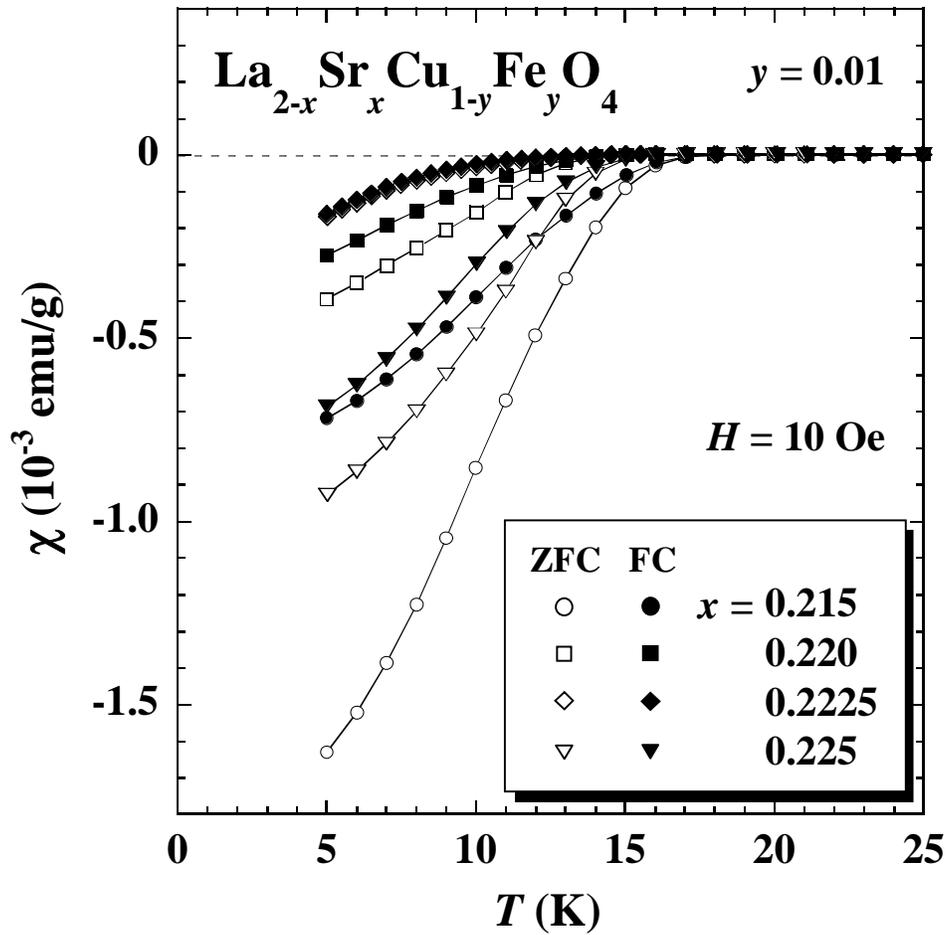}
\end{center}
\caption{Temperature dependences of the magnetic susceptibility $\chi$ in a magnetic field of 10 Oe for La$_{2-x}$Sr$_x$Cu$_{1-y}$Fe$_y$O$_4$ with $y=0.01$ on zero-field cooling (ZFC: open symbols) and field cooling (FC: closed symbols).}
\label{fig2} 
\end{figure}

\begin{figure}[tbp]
\begin{center}
\includegraphics[width=0.8\linewidth]{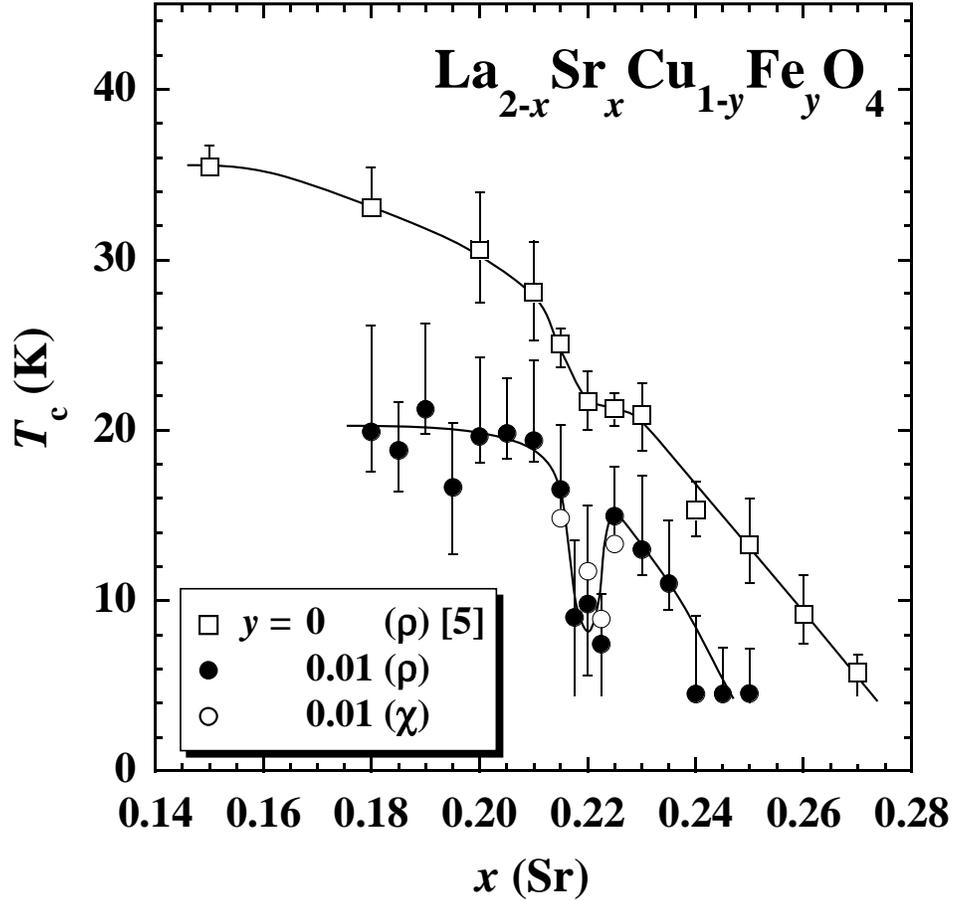}
\end{center}
\caption{Sr concentration $x$ dependence of the superconducting (SC) transition temperature $T_{\rm c}$ estimated from the electrical resistivity (closed circles) and magnetic susceptibility (open circles) measurements for La$_{2-x}$Sr$_x$Cu$_{1-y}$Fe$_y$O$_4$ with $y=0.01$. The data of $y=0$ estimated from the resistivity (open squares) are also plotted for comparison.~\cite{kaki} $T_{\rm c}$ from the resistivity measurements is defined as the midpoint of the SC transition curve. $T_{\rm c}$ from the susceptibility measurements is defined as the cross point between the extrapolated line of the steepest part of the Meissner diamagnetism and zero susceptibility. Bars indicate the temperatures where the resistivity drops to 90 \% and 10 \% of the normal-state resistivity. Solid lines are visual guides.}
\label{fig3} 
\end{figure}

\end{document}